\documentclass[%
 preprint,
superscriptaddress,
 amsmath,amssymb,
 aps,
pra,
]{revtex4-1}

\usepackage{graphicx}% Include figure files
\usepackage{dcolumn}% Align table columns on decimal point
\usepackage{bm}% bold math
\newcommand{\pddiff}[2]{\frac{\partial^2 #1}{\partial #2^2}} 
\newcommand{\ket}[1]{\left| #1 \right>} % for Dirac bras
\newcommand{\bra}[1]{\left< #1 \right|} % for Dirac kets
\newcommand{\braket}[2]{\left< #1 \, \vphantom{#2} \right|
 \left. #2 \vphantom{#1} \right>} % for Dirac brackets
\let\baraccent=\= % rename builtin command \= to \baraccent
\renewcommand{\=}[1]{\stackrel{#1}{=}} % for putting numbers above =

\usepackage{dsfont}
\usepackage[T1]{fontenc}
\usepackage[percent]{overpic}	% write text over pictures
\usepackage{color}		% adjust writing color
\usepackage{rotating}
\usepackage{placeins} % for: \FloatBarrier
\usepackage[colorlinks=true]{hyperref}
\hypersetup{%
	citecolor = blue,%
	filecolor = blue,%
	linkcolor = blue,%
	urlcolor = blue,%
	menucolor = blue%
}%

\begin{document}

\title{Quantum breathing dynamics of ultracold bosons in 1D-harmonic traps: Unraveling the pathway from few- to many-body systems}
\author{R\"udiger Schmitz}
	\email{rschmitz@physnet.uni-hamburg.de}
	\affiliation{Zentrum f\"ur Optische Quantentechnologien, Universit\"at Hamburg, Luruper Chaussee 149, 22761 Hamburg, Germany}
\author{Sven Kr\"onke}
	\email{skroenke@physnet.uni-hamburg.de}
	\affiliation{Zentrum f\"ur Optische Quantentechnologien, Universit\"at Hamburg, Luruper Chaussee 149, 22761 Hamburg, Germany}
\author{Lushuai Cao}
	\email{lcao@physnet.uni-hamburg.de}
	\affiliation{Zentrum f\"ur Optische Quantentechnologien, Universit\"at Hamburg, Luruper Chaussee 149, 22761 Hamburg, Germany}
	\affiliation{The Hamburg Centre for Ultrafast Imaging, Universit\"at Hamburg, Luruper Chaussee 149, 22761 Hamburg, Germany}
\author{Peter Schmelcher}
	\email{pschmelc@physnet.uni-hamburg.de}
	\affiliation{Zentrum f\"ur Optische Quantentechnologien, Universit\"at Hamburg, Luruper Chaussee 149, 22761 Hamburg, Germany}
	\affiliation{The Hamburg Centre for Ultrafast Imaging, Universit\"at Hamburg, Luruper Chaussee 149, 22761 Hamburg, Germany}

% \date{\today}

\begin{abstract}
Following a `bottom-up approach' in understanding many-particle effects and dynamics we provide a systematic {\it ab initio} study of the dependence of the breathing dynamics of ultracold bosons in a 1D harmonic trap on the number of bosons ranging from few to many. To this end, we employ the Multi-Layer Multi-Configuration Time-Dependent Hartree method for Bosons (ML-MCTDHB) which has been developed very recently
[S. Kr\"onke, L. Cao, O. Vendrell and P. Schmelcher. {\it New J. Phys.} {\bf 15}, 063018 (2013)].
The beating behavior for two bosons is found numerically and consequently explained by an analytical approach. Drawing on this, we show how to compute the complete breathing mode spectrum in this case.
We examine how the two-mode breathing behavior of two bosons evolves to the single-frequency behavior of the many-particle limit when adding more particles.
In the limit of many particles, we numerically study the dependence of the breathing mode frequency on both the interaction strength as well as on the particle number. 
We provide an estimate for the parameter region where the mean-field description provides a valid approximation.
\end{abstract}

\maketitle

\section{Introduction}

The boost of the physics of ultracold quantum gases has for a large part been stimulated by the experimental realization of Bose-Einstein Condensation \cite{WiemanCornell, Ketterle}.
An outstanding feature of these systems is the unique controlability, i.e. the decisive control over the external trapping
and even very fundamental properties like the interaction strength \cite{A5, A62}.
By now, also the dimensionality of the traps can routinely be tuned experimentally \cite{A32}. 
Collective oscillations have been a topic of interest right from the beginning, when they have been studied in three-dimensional trapped Bose-Einstein 
condensates \cite{A45} and are also a subject of recent research {\cite{A63}}.
This has raised the interest in the behavior of such modes in lower dimensions, where collective oscillations have already been excited and studied experimentally \cite{A32, A33}.

In the limit of many particles, the behavior of the one-dimensional breathing mode is well understood from the theoretical perspective. 
Modelling the excitations by a mean-field approach, the frequency of the
breathing mode (in a trap with frequency $\Omega$) can be calculated
analytically in the limiting cases of
an ideal gas ($2 \Omega$), a weakly interacting gas (Thomas-Fermi limit,
$\sqrt{3} \Omega$) and a Tonks-Girardeau gas ($2\Omega$) \cite{A6,A1,A14}.
These results have also been verified experimentally \cite{A32}.
Concerning the transition between these two opposite limits, the breathing frequency is amenable to a half-analytic approach by sum rules when a numerical solution of
the equilibrium ground state is provided \cite{A1}. 

In recent years there has been a growing interest in few-particle systems which have been realized and studied to a very 
high precision \cite{A56, A57, A58, A60}. Few atoms allow for analytic considerations as well as for 
\textit{ab initio} calculations, thus offer a very comprehensive insight into their physics, so there have been numerous 
theoretical efforts as well \cite{A61, A49}. 
Recently, there has been some interest in quantum many-body simulations of the breathing mode for both bosons as well as fermions \cite{A26, A27, A43}.
In the case of bosons subject to a contact-interaction, there even is an analytic 
solution of the two-body problem in a harmonic confinement \cite{A25}.
An understanding of the physics of few particle systems may then be transferred to larger systems to provide a grasp on the 
beyond mean-field physics of macroscopic ensembles. 
In this paper, we want to follow this approach and, in this sense, try to understand many-particle effects from a 'bottom-up 
perspective' by a systematic {\it ab initio} study of the impact of the number of bosons on the breathing dynamics.

\newpage
Among the collective oscillations, throughout this work we will focus on the monopole or breathing mode. Except for the dipolar oscillation with the trapping frequency, it is the energetically lowest lying mode. 
With its frequency being sensitive on the interaction regime, the breathing mode constitutes an important means of probing the interaction regime of a trapped quantum gas \cite{A32, A33}.
In fact, we study two crossovers: Due to the Bose-Fermi mapping \cite{A5}, infinitely repulsively interacting bosons behave like non-interacting fermions with respect to their spectrum and local observables. Therefor,
the breathing frequencies coincide in these effectively non-interacting limits. For a fixed number of particles, it is thus interesting to analyze the breathing frequency in between those limits where interesting correlation 
effects can be expected. On the other hand, when focusing on low interaction strengths and increasing the number of bosons the emergence of the celebrated mean-field limit should be observable in {\it ab initio} many-body calculations.

In this work we begin by exploring the breathing mode of two bosons in a one-dimensional harmonic confinement by means of the {\it ab initio} Multi-Layer Multi-Configuration Time-Dependent Hartree method for Bosons (ML-MCTDHB) \cite{B5a,B5b}. In the two-particle case, we find that with the arise of two dominant frequencies the breathing shows a beating.
These two frequencies are associated with the coupling of the ground and second excited state of the center-of-mass and relative motion. 
We indicate that, in the two-particle case, this behavior can exactly be explained by the analytic results from \cite{A25}. Using \cite{A25} the complete breathing mode spectrum is calculated and shown to consist of many frequencies which are accompanied by a multiple of sidebands each. 
We show how the analytically solvable two-particle case evolves to the single-frequency behavior of the many-particle limit.
Addressing the limit of many particles, we study the dependence of the breathing mode frequency on both the interaction strength as well as on the particle number. We discuss beyond mean-field effects and estimate the parameter range for which mean-field theory is applicable. 

This paper is organised as follows: 
In section \ref{sec:setup} we present our setup. We then discuss the two-body breathing and beating dynamics in section \ref{sec:two-body-breathing-and-beating-dynamics} before proceeding to more than two but still few particles in section \ref{sec:few-particles}. Eventually, we bridge to the many-particle case in section \ref{sec:many-particle-breathing-mode}. We then conclude this paper with a short discussion of our result and an outlook in the final section \ref{sec:summary-and-outlook}.

\section{Setup and method}
\label{sec:setup}

Let us consider a system of $N$ bosons of mass $m$ in a one-dimensional harmonic
trap with trapping frequency $\Omega_0$.
The breathing mode is triggered by a sudden quench of the trap frequency. 
Due to symmetry, this procedure will, independently of 
e.g. the particle number, not excite a dipolar oscillation but solely a breathing dynamics. 
Rescaling the one-dimensional harmonic oscillator Hamiltonian in harmonic oscillator units, it reads
\begin{equation}
      \hat{\tilde{H}} = -\frac{1}{2}\sum_{k=1}^N \pddiff{}{\tilde{x}_k} + \frac{1}{2} \sum_{k=1}^N \tilde{\Omega}^2 \tilde{x}_k^2 + \tilde{g} \sum_{k<l} \delta(\tilde{x}_k - \tilde{x}_l)
\end{equation}
such that lengths are given in units of the initial oscillator length $a_0=\sqrt{\frac{\hbar}{m\Omega_0}}$, energies with respect 
to $\hbar \Omega_0$ and frequencies in terms of the initial trap frequency, i.e. $\tilde{x_k} = x_k / a_0$, $\tilde{H}=H/ \hbar \Omega_0$ 
and $\tilde{\Omega} = \Omega / \Omega_0$.
The interaction strength is scaled as 
$\tilde{g} = \frac{g_{\rm{1D}}}{\hbar \Omega_0 a_0}$.
In the following, we will drop the tilde as we are only concerned with the rescaled units. 

Note that $\langle \hat X^2 \rangle$ relates to the expectation value of a one-body observable in fact, namely 
$\langle \hat X^2 \rangle = N \text{tr} (\hat x^2\hat \rho_1)$ with $\hat \rho_1$ denoting the reduced one-body 
density operator. If only the monopole mode is excited, $\text{tr} (\hat x\hat \rho_1)$ vanishes and then 
$\langle \hat X^2 \rangle$ becomes a measure for the variance of the single particle density.
Due to the discreteness of this trapped quantum system, separated peaks arise in the Fourier spectrum for sufficiently long propagation times. We determine the peak positions either by locally fitting Lorentzians to the peaks or - if we can be sure that only a single frequency dominantly contributes to the spectrum - by fitting a sine function to the real time data.

The truncation of the many-body Hilbert space to a variationally optimal, time-dependent subspace makes the propagation of such systems and 
the study of beyond-mean-field effects feasible. Such a scheme is employed by the family of Multi-Configuration
Time-Dependent Hartree (MCTDH) methods \cite{B1,B2}. 
With Multi Layer-MCTDHB, a powerful generalization has been developed very recently for applications to the quantum dynamics of ultracold bosonic pure and multi-species systems \cite{B5a,B5b}.
A brief discussion of this method can be found in Appendix \ref{sec:Appendix_MLB}.
At this point, we just want to note that we are able to propagate the system using an \textit{ab initio} method considering 
all correlations. The control parameter of all MCTDH methods is the number of orbitals provided (i.e. the dimension 
of the subspace the Hilbert space is truncated to), which we denote by $M$. Convergence is ensured if increasing $M$ does not 
change the observables of interest and in this case all relevant correlations are 
taken into account. 

If not stated otherwise, all the following data are obtained by exciting the breathing mode through a quench of the trapping frequency from $\Omega = 1.0$ to $\Omega = \sqrt{0.9}$.

\section{Two-body breathing and beating dynamics}
\label{sec:two-body-breathing-and-beating-dynamics}

\begin{figure}[b]
   \centering %
   \begin{overpic}[width=0.5\linewidth, keepaspectratio]{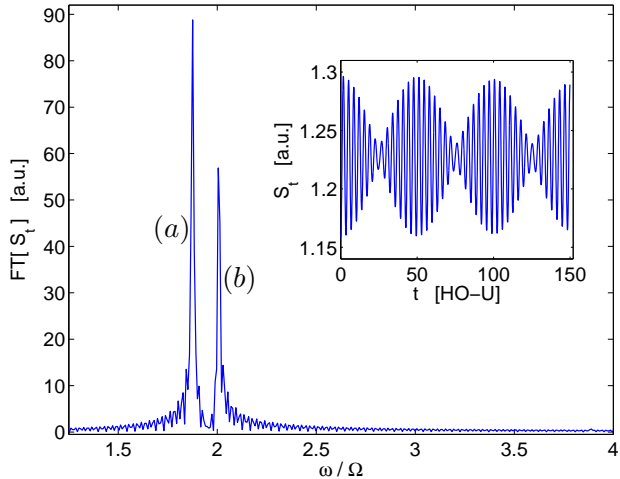}%
	 \put(24,40){\textcolor{black}{$(a)$}}%
	 \put(35.2,32){\textcolor{black}{$(b)$}}%
   \end{overpic}
 \caption[Ocurence of a beating in the breathing mode and Fourier spectrum of the breathing mode]{Occurence of a beating in the breathing mode as indicated by $\langle \hat{X}^2 \rangle$ (inset) and its Fourier spectrum.
   The signal $S_t$ refers to this expectation value, $S_t = \langle \hat{X}^2 \rangle$. Two orbitals have been provided, 
   the breathing has been triggered by quenching the trap from $\Omega=1$ to $\Omega=\sqrt{0.9}$. [HO-U] refers to harmonic oscillator units with respect to the Hamiltonian before the quench.}%
 \label{FigFirstBeating}
 \end{figure}

Following our `bottom-up approach', let us restrict ourselves to the simplest nontrivial case, i.e. two bosons.
In figure \ref{FigFirstBeating} we depict typical results on a two particle breathing, showing $\langle \hat{X}^2 \rangle$
as a function of $t$. 
The system not only features a breathing but also a beating is superimposed (figure \ref{FigFirstBeating}, inset), which corresponds to two clearly dominant frequencies in the $\langle \hat{X}^2 \rangle$ spectrum. 

First, we numerically examine this two-mode dynamics at different interaction strengths and with different orbital numbers. 
Whilst the higher-frequency peak (indicated by (b) in figure \ref{FigFirstBeating}) happens to be insensitive to varying the interaction 
strength $g$, given that sufficiently many orbitals are provided (cf. appendix \ref{sec:Appendix_Separation}), the lower-frequency 
peak (indicated by (a) in figure \ref{FigFirstBeating}) is strongly affected by $g$.

%[ Figure: 4.3 with 4.2 as an inset ]

%[ Figure 4.9 ]
 \begin{figure}[t]
   \centering %
     \includegraphics[width=0.8\linewidth, keepaspectratio]{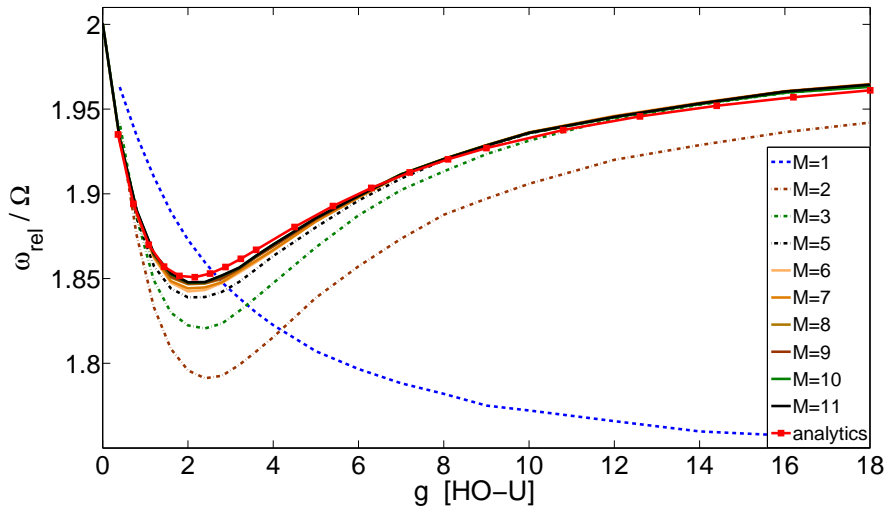}%
   \caption[Relative motion breathing mode frequency divided by trap frequency for two particles and $M=1,\ldots,11$ orbitals as functions of $g$ and theory.]{(Color online) Relative motion breathing mode frequency for two 
   particles and $M=1,\ldots,11$ orbitals as functions of $g$. The error owing to the finite spectral resolution is $\Delta \omega = 0.013$.}%
   \label{FigRun39AlleMitTheorie} %
 \end{figure}

In figure \ref{FigRun39AlleMitTheorie} we show our numerical results for the dependence of the lower-frequency peak on the 
interaction strength: 
For zero interaction, the breathing frequency divided by the frequency of the trap equals two. The same value is 
approached for a very strong repulsion. In between, there is a local minimum at $g\approx 2$. 
Compared to the theoretical results we will derive later on, 
we see a very good agreement for higher orbital numbers. 
Further, our analysis \footnote{We note that, independently of our above analysis, there has already been numerical evidence 
on the occurrence of a beating in the breathing mode \cite{A26, A27}. 
The authors interpreted the two peaks as arising from the mutually decoupled relative and center-of-mass motion \cite{A43}.
Assuming Coulomb interaction, however, the system has 
not been amenable to a comprehensive exact analysis \cite{A26, A27, A43}. A related approach has already been applied to a 
different setup, namely a two-boson system driven by a time-dependent interaction \cite{A51}.} 
will show that this peak can utterly be explained by the relative motion of the two-particle system.

From the results in figure \ref{FigRun39AlleMitTheorie} it is evident that mean-field calculations ($M=1$) cannot 
account for the behavior 
of the lower frequency when approaching fermionization (i.e. the high-$g$ limit). For the local observable $\hat X^2$, 
the $g=0$ and $g=\infty$ limit represent effectively interaction-free cases. Therefore, there should be some finite $g_m>0$ 
for which the interaction effects are most dominant. In the mean-field picture, however, there is no finite $g$ at which the 
interaction effects are maximal. Rather than returning asymptotically to the non-interacting value, the mean-field solution 
monotonically approaches the Thomas-Fermi limit $\sqrt{3}\Omega$, i.e. the strong interaction limit within the mean-field theory, 
and is thus not capable of resolving the reduction of interaction effects on $\omega_{\rm{rel}}$ when approaching the Tonks Girardeau 
limit. Moreover, the mean-field description does not even reveal that there is any beating at all. 
The occurrence of such a 'beating' mode thus is a pure beyond-mean-field effect, which is not unexpected to occur for a few-boson system.

Let us now try to understand the physical reason why this simple system features such a rich breathing spectrum.
Due to the harmonic trapping, we may separate the Hamiltonian as
\begin{equation}
	\hat{H} = \hat{H}_r + \hat{H}_R\, ,
	\label{EqSeparationHamiltonian}
\end{equation}
with
\begin{align}
	\nonumber
	\hat{H}_R & = -\frac{1}{2} \pddiff{}{R} + \frac{1}{2} \Omega^2 R^2\, ,
	\\
	\nonumber
	\hat{H}_r & = -\frac{1}{2} \pddiff{}{r} + \frac{1}{2} \Omega^2 r^2 + g \delta (r) \, ,
\end{align}
where we have introduced the center-of-mass (CM) coordinate ${R = \frac{1}{\sqrt{2}}(x_1 + x_2)}$ and the relative coordinate 
$r = {\frac{1}{\sqrt{2}}(x_1 - x_2)}$ \footnote{Please note that as in \cite{A25} we employ the symmetric definition of the CM and relative coordinate leading to an equal total and reduced mass.}.
The relative motion part $\hat{H}_{\rm{r}}$ depends on the interaction strength $g$, whereas the CM motion part does not. 
For both parts we recover a harmonic oscillator in $r$ and $R$, respectively, whilst the former is superimposed by a 
delta interaction $g\delta(r)$.
In the following, let us denote all properties belonging to the CM motion by capital letters and use lowercases 
for the relative motion. 

Stepping back to only one particle in a harmonic trap for a moment, we can write 
$\hat x^2 = \frac{1}{2\Omega} (\hat a^\dagger + \hat a)^2$, where 
$\hat{a}^\dagger$ and $\hat{a}$ denote the usual single-particle harmonic oscillator creation and annihilation operators 
obeying $[\hat{a},\hat{a}^\dagger ] = \mathds{1}$. 
Thus we see that in the single-particle case the operator $\hat{x}^2$ can couple states of at most two quanta difference, which 
moreover have to have the same parity. In the two-particle case, this will be exactly the same for the CM motion 
part, i.e. $\hat{R}^2$, since the CM motion Hamiltonian of the interacting two-particle case corresponds to just a 
pure single-particle harmonic oscillator. 
To the relative motion part $\langle \hat{r}^2 \rangle$, however, also states with quantum numbers differing by other than two 
may contribute, since the relative motion part is not a pure harmonic 
oscillator anymore for a non-vanishing interaction strength. 

\begin{figure}[htbp]
   \centering %
   \includegraphics[width=0.6\linewidth, keepaspectratio]{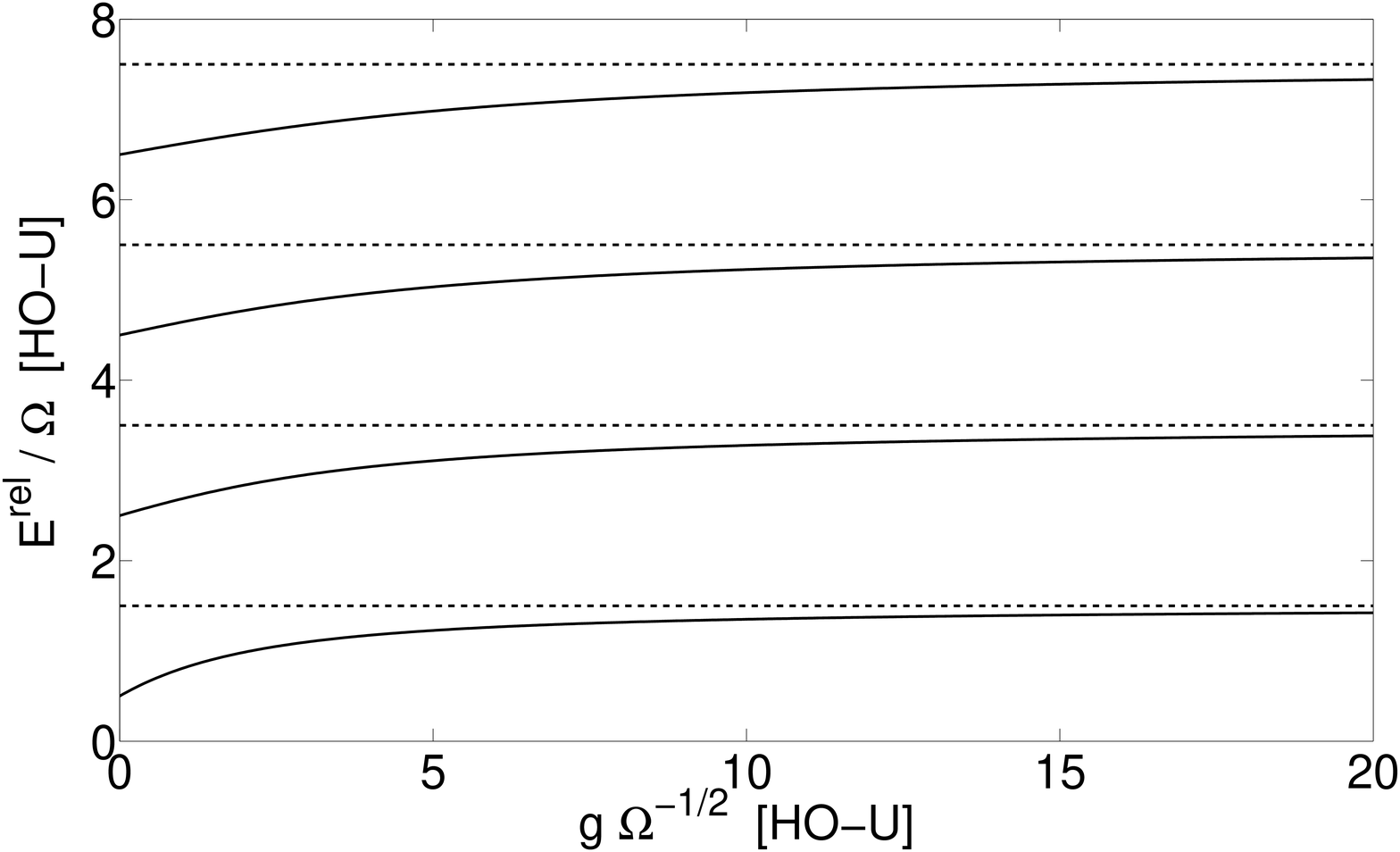}%
   \caption[Energy of the relative motion for two particles with an arbitrary interaction strength confined by a Harmonic trap. Cf. \cite{A25}.]{Energy of the relative harmonic oscillator for two particles with an arbitrary interaction strength. 
   The broken lines indicate the asymptotic values that are reached by the energies (solid lines) in the limit $g\rightarrow \infty$.
   Computed using the analytic solution in ref. \cite{A25}. Cf. figures therein.}%
   \label{FigBuschEnergies} %
 \end{figure}

The behavior of the relative motion can be understood in detail since there exists an analytic solution for the energy spectrum 
of the relative motion harmonic oscillator for two bosons in a harmonic trap, subject to a contact interaction \cite{A25}.
Figure \ref{FigBuschEnergies} shows the energy spectrum of the relative motion Hamiltonian $H_r$ depending on the interaction 
strength $g$, which is computed using the analytic solution given in ref. \cite{A25}.
% Figure 4.4

Let us expand the initial state with respect to the eigenstates of the Hamiltonian after the quench,
\begin{align}
	\ket{\psi} = \sum_{i,I} \alpha_{2I,2i} \ket{\phi_{2i}}_{\rm{r}} \otimes \ket{\Phi_{2I}}_{\rm{R}} \, ,
\end{align}
where $\alpha_{2I,2i}$ denotes the contribution of the respective basis state. For the relative motion part, only even 
states contribute because of the bosonic exchange symmetry. Due to the parity symmetry, the same holds for the center-of-mass 
part. Furthermore, the time-reversal symmetry of the one-dimensional Hamiltonians $\hat H_r$ and $\hat H_R$ (before and after the 
quench) allows us to assume the respective eigenfunctions and, thus, the coefficients $\alpha_{2I,2i}$ to be real. Bearing in mind that $\hat{R}^2$ only couples 
eigenstates of $\hat H_R$ with quantum numbers differing by two, whereas the 
operator $\hat{r}^2$ can in principle couple any pair ${(\,\ket{\phi_{2i}},\ket{\phi_{2j}}\,)}$, we find that
\begin{align}
	\nonumber
	\langle \hat{X}^2 \rangle & = \langle \hat{R}^2+\hat{r}^2 \rangle 
	\\
	\nonumber
	& = const.  + 2\sum_{I, i=0}^\infty F_{2I,2(I+1)} \,\alpha_{2I,2i} \,\alpha_{2(I+1),2i} \, \times 
	\\
	\nonumber
	& \qquad \quad \times \cos \left\{ 2 \Omega t \right\} + 
	\\ 
	\nonumber
	 & \quad + 2 \sum_{i>j=0, I= 0}^\infty f_{2i,2j}  \,\alpha_{2I, 2i}
	 \, \alpha_{2I, 2j} \, \times 
	 \\
	 & \qquad \quad \times \cos \left\{ \Omega \, [ (2i-2j) - \Delta_{2i,2j}(g) ] \, t \right\} \, ,
	\label{EqBeating}
\end{align}
where the first sum comes from the CM motion harmonic oscillator and couplings of the form 
$\bra{\Phi_{2I};t} \hat{R}^2 \ket{\Phi_{2J};t}\cdot \braket{\phi_{2i}}{\phi_{2i}}$, whereas the second one corresponds to 
$\bra{\phi_{2i};t} \hat{r}^2 \ket{\phi_{2j};t}\cdot \braket{\Phi_{2I}}{\Phi_{2I}}$ couplings of the relative motion.
The functions $F$ and $f$ denote the respective matrix elements at time zero, i.e. 
$F_{2I,2J}=\bra{\Phi_{2I};t=0} \hat{R}^2 \ket{\Phi_{2J};t=0}$ and $f_{2i,2j}=\bra{\phi_{2i};t=0} \hat{r}^2 \ket{\phi_{2j};t=0}$.
In the last term, we have introduced $Delta_{2i,2j}(g) := [\epsilon_{2j}(g) - \epsilon_{2i}(g)] / \Omega$
with $\epsilon_{2j}(g)$ referring to the energy shift of $\ket{\phi_{2j}}$ with respect to the non-interacting case, i.e.
${\epsilon_{2j}(g)=E_{2j}^{\,\rm{rel}}(g) - E_{2j}^{\,\rm{rel}}(0)}$.

Let us figure out which of these frequencies are responsible for the beating behavior we addressed previously. For a weak quench, 
we might expect that the lowest lying states are dominant. This means that for both relative and center-of-mass motion, 
the respective ground state and the second excited state are the most important ones. According to (\ref{EqBeating}) this gives rise to two frequencies:

(i) One from the center-of-mass coupling, $F_{0,2} \alpha_{0,2i} \,\alpha_{2,2i} \,\cos \left\{ 2 \Omega t \right\}$, at $2\Omega$. 
As it comes from matrix elements of the form $\bra{\Phi_0;t} \hat{R}^2 \ket{\Phi_2;t} $ we call it the center-of-mass motion breathing 
frequency. Apparently, its frequency is independent of the interaction strength $g$. It corresponds to the peak (b) in 
figure \ref{FigFirstBeating}. \\
Indeed, for a sufficiently high orbital number, we have found that this peak remains 
almost constant when varying the interaction strength. For a numerical subtlety of MCTDH-type methods regarding the separation of 
center-of-mass and relative motion, we refer to appendix \ref{sec:Appendix_Separation}.

(ii) Another frequency, peak (a) in figure \ref{FigFirstBeating}, comes from the last sum, namely from terms of the form
$f_{2,0} \alpha_{I, 0}\, \alpha_{I, 2} \, \cos \left\{ \Omega [ 2 - \Delta_{2,0}(g) ] \right\}$. These arise from the 
matrix elements $\bra{\phi_0;t} \hat{r}^2 \ket{\phi_2;t} $, thus this peak can be assigned to the relative motion. It 
lies just below $2\Omega$, shifted to a lower frequency by $\Delta_{2,0}(g)\geq 0$. This shift comes from the fact that the ground 
state and the second excited state of the relative motion are subject to different energy shifts (cf. figure \ref{FigBuschEnergies}). 
This relative motion breathing frequency is $g$-dependent through $\Delta_{2,0}(g) $. 
As apparent from figure \ref{FigBuschEnergies}, the energy levels at infinite (repulsive) interaction strength will agree with the 
non-interacting ones, which can also be inferred from the Bose-Fermi mapping \cite{A6}. Thus, for zero as well as for infinite interaction strength, there is no 
beating present and the breathing mode will consist of one single frequency at $2\Omega$. 
Calculating $\Omega [ 2 - \Delta_{2,0}(g) ]$ leads to the red curve in figure \ref{FigRun39AlleMitTheorie} which shows a good
agreement with the numerical results for a sufficiently high orbital number. 

As not only the second excited state and the ground state of the relative motion degree of freedom are differently affected by 
(particularly low) interactions but indeed \textit{every} state is subject to yet another energy shift, each higher 
excitation will cause new frequencies to arise. 
All those pairs of quantum numbers differing by two will cause other frequencies slightly below $2\Omega$. Thus, from the set 
of all such pairs, a band of frequencies just below 2 arises.
The same argument applies to pairs of states differing by $4$, $6$, $\ldots$ in their quantum numbers, thus forming such bands 
around $4\Omega$, $6\Omega$, $\ldots$ as well. 

\begin{figure}[t]	%change SK
   \centering %
   \includegraphics[width=1.0\linewidth, keepaspectratio]{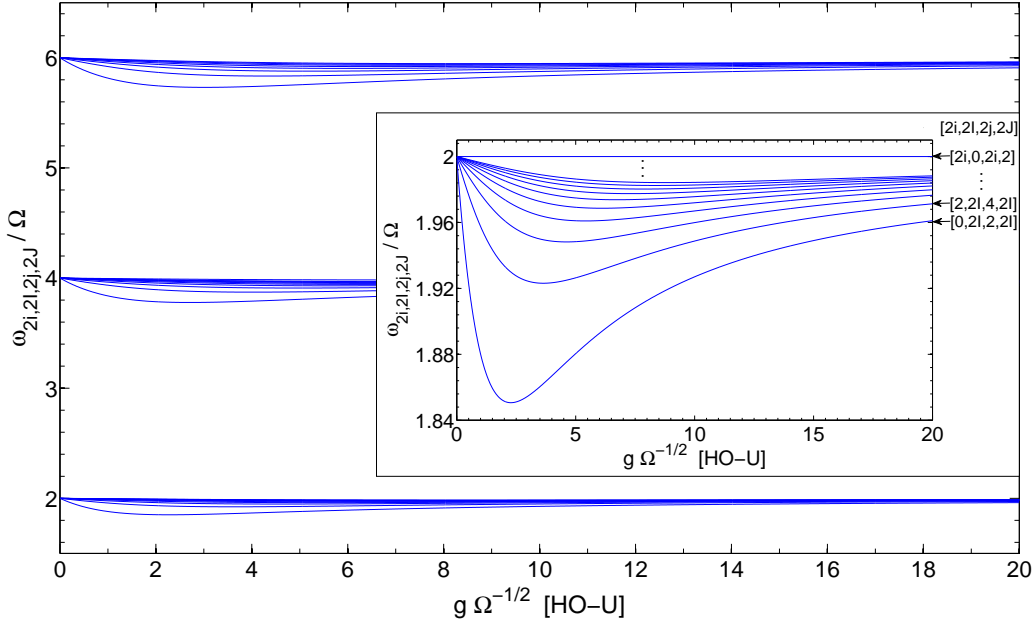}%
  \caption[Prediction for the full breathing/ beating mode spectrum at any interaction strength.]{Prediction for the full breathing mode spectrum at any interaction strength, up to $\approx 6\Omega$.
     Here, we have depicted all such modes from states with up to 20 quanta. The inset provides a detailed view on the lowest band.
     The frequencies are labelled by $\omega_{2i,2I,2j,2J}$ which refers to the frequency arising from $\bra{\Phi_{2I}\phi_{2i}} \hat{X}^2 \ket{\Phi_{2J}\phi_{2j}}$.}%
   \label{FigFullSpectrum} %
 \end{figure}

\begin{figure}[htb]
   \centering %
   \includegraphics[width=0.7\linewidth,keepaspectratio]{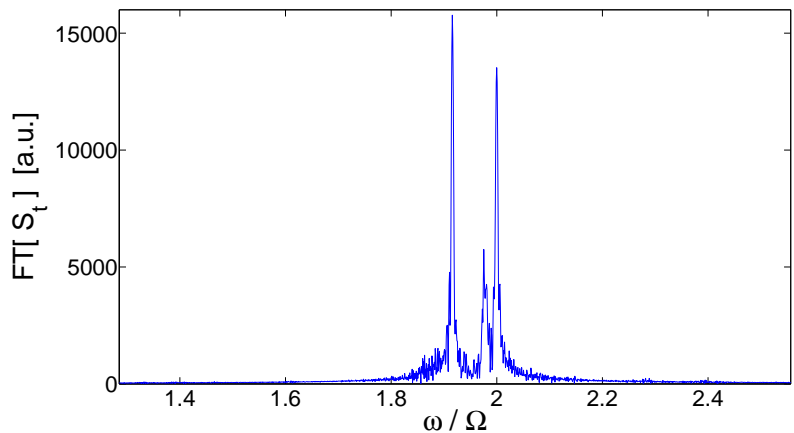}%
   \caption[Numerical hint on the existence of sidebands]{Numerical hint on the existence of sidebands. The signal $S_t$ refers to the expectation value of $\hat{X^2}$, $S_t = \langle \hat{X}^2 \rangle$. 
   Note that in this case the trap has been quenched to $\Omega=\sqrt{0.3}$.
   The interaction strength has been set to $g=0.4$. The peak positions are $\omega_1/\Omega=1.916$, $\omega_2/\Omega=1.975$, $\omega_3/\Omega=2.000$.
  The position of the middle peak $\omega_2$ agrees with the first sideband $\omega_{2,2I,4,2I}$ in figure \ref{FigFullSpectrum}, coming from $\bra{\phi_2}\hat{r}^2\ket{\phi_4}$. }%
   \label{FigSideBand}
 \end{figure}

The above analysis results in the spectrum depicted in figure \ref{FigFullSpectrum}. The inset gives a detail view on the lowest lying band (where the lowest lying curve corresponds to the 
analytically obtained, red curve in figure \ref{FigRun39AlleMitTheorie}).
The other bands are similar except for the fact that they lack a center-of-mass motion peak. 
%Figure 4.8 with 4.6 as an inset
In accordance with our analysis, around $2\Omega$ we have found numerical evidence of the existence of sidebands as indicated by 
figure \ref{FigSideBand}. In order to amplify the sidebands, which stem from higher excitations of the relative motion harmonic 
oscillator, we have quenched the trap from $\Omega=1$ to $\sqrt{0.3}$.

Concerning the experimental realizability we stress that most of the above modes, particularly all the sidebands, are strongly suppressed by the weak occupation of the contributing 
states. 
Obviously, the dominant relative motion frequency of the $\ket{\phi_2}$-$\ket{\phi_0}$-coupling and the center-of-mass mode are the easiest to access experimentally. 
The key issue is that - like for our numerics as well - one has to ensure a sufficiently long propagation time to resolve the two frequencies and their spectral separation. As deducible from the inset of figure \ref{FigFullSpectrum}, we expect the separation to be approximately $7.5\,\%$ of the total breathing frequency.
In \cite{A32} a breathing frequency of around $150\, Hz$ is reported. 
With the spectral separation of the two breathing frequencies being of the order of some percents ($<7.5\, \%$), one cycle of the beating, oscillating with half the frequency separation, would thus last $\approx 0.2 s$ in this setup, which enlightens the experimental challenge.

\section{Few-particle breathing mode}
\label{sec:few-particles}

With the understanding of the two-particle case we now want to approach the regime of few particles. To this end, we need to study 
how our findings in the two-particle case compare with the theory outlined in the introduction. 
In particular, for an increasing
particle number the center-of-mass breathing mode is strongly suppressed and breathing of the relative mode becomes dominant, i.e. one obtains effectively a single frequency of breathing, and 
$\omega_{\rm{br}}/\Omega$
should approach the mean-field Thomas Fermi value of $\sqrt{3}$ for weak but dominant interactions.

First of all, let us discuss how our findings of the two-particle many-mode breathing/ beating reduce to the well-known 
breathing behavior for increasing particle number. 
The answer to this issue can be found in \cite{A43} where the authors have recently reported on the occurrence and behavior of 
two frequencies in Coulomb interacting particle systems.
Casting the $N$-particle Hamiltonian in coordinates of CM and relative motion, $R = \frac{1}{N} \sum_{i=1}^N x_i$ and 
$r_i = x_i - x_{i-1}$, we find that the CM coordinate obeys a harmonic oscillator equation with mass $N$, the Hamiltonian reads $\hat{H}_R = - \frac{1}{2N} \pddiff{}{R} + \frac{1}{2} \Omega^2 N R^2 $. Thus, the breathing 
mode amplitude of the CM motion $\langle \hat{R}^2 \rangle$ is suppressed, as by rescaling $R^\prime = \sqrt{N} R$ we find that $ \langle \hat{R}^{2} \rangle = \frac{1}{N} 
\langle \hat{R}^{\prime 2} \rangle$ \footnote{This is not 
in agreement with \cite{A43} where a factor of $1/\sqrt{N}$ is given.}. 

\begin{figure}[t]
   \centering 
       \includegraphics[width=0.7\linewidth,keepaspectratio]{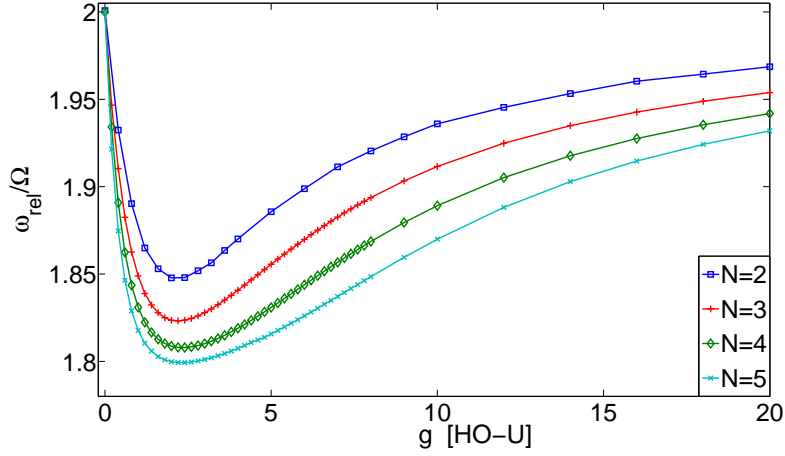}%
    \caption[Relative motion peak for $3$, $4$ and $5$ particles at different orbital numbers as a function of $g$]{(Color online) Breathing frequency of the relative motion for $3$ 
  (red, plus markers), $4$ (green,  diamonds) and $5$ (light blue, crosses) particles
     as a function of $g$. For all these cases, $M=9$ orbitals have been provided. For comparison, we have also depicted the graph for two particles and $M=11$ (dark blue, squares).}%
   \label{Fig3-4-5Particles} %
 \end{figure}

As in the previous section, we study the breathing frequency of the relative motion. The results for $3$ to $5$ particles are given in figure 
\ref{Fig3-4-5Particles} where we have also included the two-particle behavior for comparison. We see that with increasing particle number, 
the breathing frequency of the relative motion exhibits a deeper minimum, which lies above the value of the Thomas Fermi prediction $\omega_{\rm rel} / \Omega = \sqrt{3}$ for the particle numbers under consideration.
Qualitatively, however, the curves for higher particles numbers $3$ to $5$ show the same behavior as for two particles.

\section{Many-particle breathing mode}
\label{sec:many-particle-breathing-mode}

Having learned how the beating excitation transforms into a single breathing mode with increasing particle number we now want to explore the 
behavior of the breathing mode 
up to yet higher particle numbers. Due to computational limitations, the feasible number of orbitals is quite restricted. 
This means a restriction on the extent to which correlations can be considered and consequently we have to focus on the low interaction regime.

\begin{figure}[b]
   \centering %
     \includegraphics[width=0.9\linewidth,keepaspectratio]{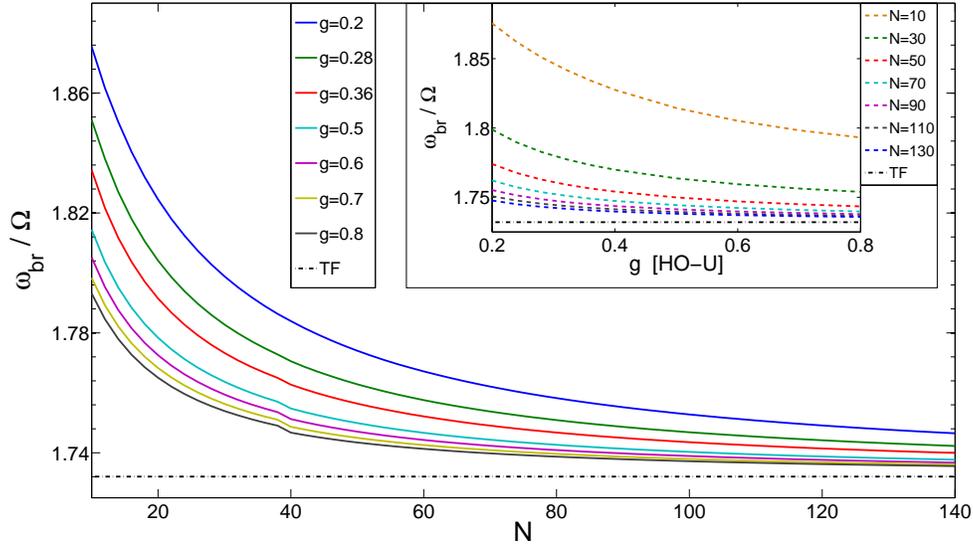}%
   \caption[Breathing mode frequency as a function of interaction strength and
 particle number]
 {(Color online) Breathing mode frequency as a function of 
  particle number for various interaction strengths ranging from $g=0.2$ (uppermost) to
  $g=0.8$ (lowermost solid line). Inset: Breathing mode frequency as a function of 
  interaction strength for various particle numbers ranging from $N=10$ (uppermost) to
  $N=130$ (lowermost dashed line). Dashed dotted lines: breathing frequency within 
  Thomas Fermi approximation.} 
  \label{Fig-N-g} %
 \end{figure}

First, let us, for a fixed particle number, vary the interaction strength just as we did in the preceding section. We have 
depicted the results for different particle numbers in the inset of figure \ref{Fig-N-g}.
For a higher particle number, but fixed $g$,
the system features a lower-lying breathing frequency. Thus, judging from the breathing mode, for a many-particle system 
the transition from the ideal to the weakly-interacting gas, i.e. the
mean-field limit, happens 'faster' with respect to the interaction strength.

\FloatBarrier

Complementary to the above, let us now consider the breathing mode frequency as a function of the particle number, i.e. for a fixed interaction strength.
This is indicated by figure \ref{Fig-N-g} for a number of different interaction strengths. 
We see that for increasing particle number the breathing mode approaches the results from the mean-field limit and this happens the faster the higher the interaction strength is.
Note that we hereby exclusively consider the case of weak interactions. Contrary to the results of \cite{A43} for fermions, the breathing frequency as a function
of the particle number seems not to exhibit a minimum for this regime.
The minor step between $N=38$ and $N=40$ for $g \leq 0.5$
stems from the fact that computational restrictions have forced us to reduce the number of orbitals from $M=4$ for $N\leq 38$ to 
$M=3$ for $38<N\leq 140$. Setting $M=4$ for $N\geq 40$ would result in a too excessive number of coefficients for long time propagations (cf. appendix \ref{sec:Appendix_MLB}). 
Please note that for any $M$, the ML-MCTDHB algorithm provides the variationally optimal approximation to 
the true many-body wavefunction. The reduction of $M$ here just leads to a small quantitative deviation from the expected behavior.

\begin{figure}[b]
   \centering %
     \includegraphics[width=0.95\linewidth, keepaspectratio]{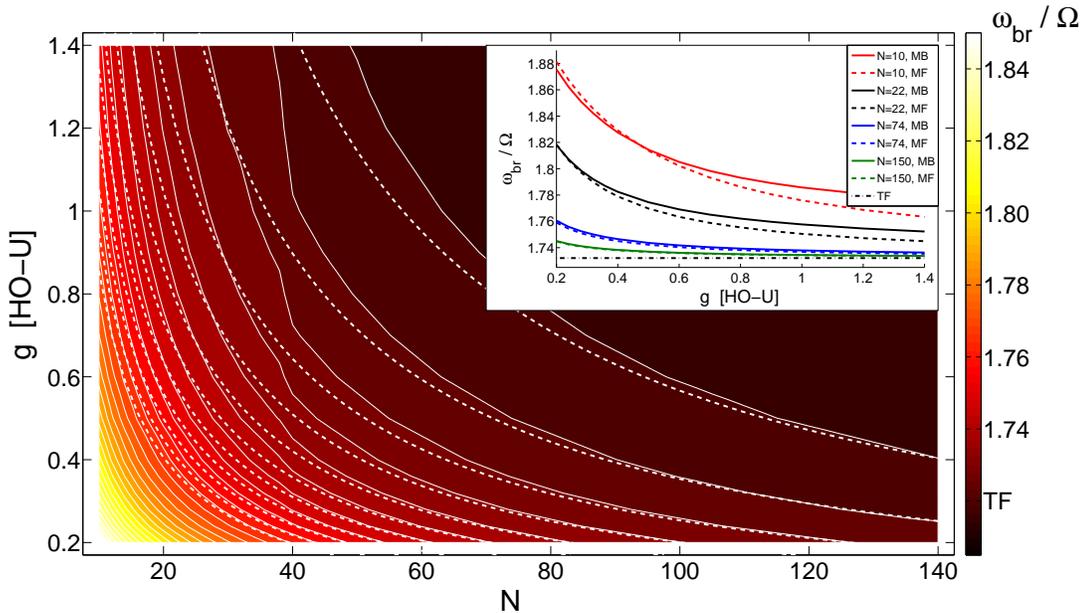}%
   \caption[Contour plot of the breathing mode frequency, fit with curves of constant Gross-Pitaevskii parameter]{(Color online) Contour plot for the frequency of the breathing mode $\omega_{\rm{br}}/\Omega$. The white dashed lines
   indicate lines of constant $g\cdot(N-1)$. The inset shows the dependence of the breathing frequency on the interaction strength for some characteristic particle numbers and compares these curves with the corresponding mean-field results.
   `MF' denotes mean-field, i.e. effective one-body results, whereas `MB' refers to many-body, i.e. converged ML-MCTDHB results.}%
   \label{FigContourFit} 
 \end{figure}

In figure \ref{FigContourFit} (inset) we compare the many-body results for the breathing mode frequency as a function of the 
interaction strength for $N=10$, $22$, $74$, $150$ with the corresponding mean-field results.
The mean-field calculations accounts for the transition from the ideal gas to the Thomas-Fermi regime. 
Generally, they tend to overestimate the breathing frequency for very low $g$, as deducible from the $N=10$ curve. The same
behavior can be seen for higher particle numbers if we continue these curves to lower values of $g$ (not provided in the inset). 
The mean-field calculations cannot show any features of the fermionization limit, hence it is obvious that for higher interaction
strengths the mean-field results underestimate the many-body results, since the latter are affected by the breathing mode frequency surge 
when approaching fermionization whilst the former are not. 
Continuity implies that there is a point where the many-body and the mean-field curves intersect each other.
Away from these intersection points, it is apparent that the mean-field results become the better the higher the particle number is.

As already deducible from figure \ref{Fig-N-g} (compare main figure and inset) there is a symmetry of the system with respect to particle number and 
interaction strength. Both parameters influence the breathing mode frequency in a similar way. 
This is expected as for low interaction strengths the system is governed by the Gross-Pitaevskii parameter $g\cdot (N-1)$. 
Note that one can show that the Gross-Pitaevskii equation is exact in the limit $g \rightarrow 0$, $N \rightarrow \infty$ with $g(N-1)=const.$ \cite{A55,A64}.
Conversely, we may use this fact to study the parameter range in the $g-N$ plane in which mean-field theory is well applicable.
Since in this limit the properties of the gas are given by the Gross-Pitaevskii parameter alone, the breathing frequency will remain constant for constant
$g\cdot (N-1)$ as long as Gross-Pitaevskii theory is applicable.
If we compare lines of constant 
$g\cdot (N-1)$ with those of a constant breathing mode frequency, we will thus be able to estimate the region in $g$-$N$-parameter space which 
mean-field theory may properly be applied to.
In figure \ref{FigContourFit}, we have included curves of constant $g(N-1)$ such that they agree with the contour lines of our data at their respective right edge. 
We can clearly identify those regions where $g(N-1)$ proves to be a good parameter in a non-trivial surrounding and those where this is not the case. For the latter, 
we can deduce that mean-field theory is not applicable and beyond mean-field effects are important. For the former, $g(N-1)$ being a good parameter suggests that mean-field theory 
is well applicable, even though this implication is not strict. We note that for this statement to hold strictly, it would be necessary to compare the contour lines in our data 
to those hyperbolas with which they agree asymptotically. From the comparison of the contour lines and
the Gross Pitaevskii hyperbolas, one can, however, infer the extent of the regions of local agreement, which increases for larger particle numbers and lower interaction strengths. In this
sense, one may estimate the region of mean-field like breathing behaviour.
\FloatBarrier

\section{Summary and Outlook}
\label{sec:summary-and-outlook}

We have investigated the dynamics of the breathing mode following a quench of a one-dimensional harmonic trap. Our focus was the crossover from few- to many-body bosonic systems with an emphasis on the emergence of the mean-field behaviour for many bosons and weak interactions.
By state of the art methods like MCTDHB or our recently developed ML-MCTDHB, such extensive {\it ab initio} studies have become possible.
In the two-particle case, we have found that with the arise of two frequencies the breathing shows a dominant beating. 
Using the solution of the two-particle problem \cite{A25}, the complete breathing mode spectrum has been calculated. It consists of 
many frequencies which are accompanied by a multiple of sidebands each. Experimental evidence for the beating behavior in 
general remains as a challenging task. For instance on the basis 
of a breathing frequency of around $150\, Hz$, as reported in \cite{A32}, we expect 
that the observation of the beating and measuring of the two contributing frequencies would require an evolution time of the 
order of $0.1 \, s$. 
We have shown how the analytically solvable two-particle case evolves to the single-frequency behavior of the many-particle limit when adding 
more and more particles.
In the limit of many particles, we have numerically studied the dependence of the breathing mode frequency on both the interaction strength 
as well as on the particle number. In particular we have provided an estimate for the parameter region in which the Gross-Pitaevskii mean-field approach is applicable. 

One possible extension of the present work are mixtures of different bosonic species. 
For this case it would be interesting to study how a similar beating could arise 
from different breathing frequencies the constituents may obey due to different particle numbers, interaction strengths or 
trapping potentials. 
Due to its multi-layer structure, ML-MCTDHB is a most suitable {\it ab initio} method for simulation such complex bosonic systems.

\begin{acknowledgments}
The authors would like to thank Ioannis Brouzos for valuable discussions.
L.C. and P.S. gratefully acknowledge funding by the
Deutsche Forschungsgemeinschaft in the framework of the SFB 925 ``Light induced
dynamics and control of correlated quantum systems''. R.S. and S.K. gratefully acknowledge scholarships of the Studienstiftung des deutschen Volkes.
\end{acknowledgments}

\appendix
\section{The ML-MCTDHB method}
\label{sec:Appendix_MLB}

The \textit{Multi-Layer Multi-Configuration Time-Dependent Hartree method for Bosons}, ML-MCTDHB, is a variational {\it ab initio}
method for studying the non-equilibrium dynamics of bosonic systems \cite{B5a,B5b}.
The idea behind all MCTDH-type methods is to represent the wave function by a number of variationally optimal chosen orbitals, i.e. 
to truncate the Hilbert space to a variationally optimal subspace.
In order to ensure that this subspace remains optimal throughout the propagation, the orbitals are time-dependent. 

Let us briefly introduce the 
ansatz for the many-body wave function and sketch the derivation of the
equations of motions for the Multi-Configuration Time-Dependent Hartree method for Bosons (MCTDHB) \cite{B17, B4, B18} 
to which ML-MCTDHB reduces in the case of a single species
in one dimension.
We expand the wave function in terms of time-dependent \textit{permanents},
\begin{equation}
    \ket{\Psi(t)} =\sum_{n_1,...,n_M} C_{n_1,...,n_M}(t) \ket{n_1, n_2, \ldots, n_M; t}
    \label{EqMCTDHBAnsatz}
\end{equation}
where $M$ denotes the number of orbitals and the $n_i$'s sum up to the total number of bosons. The permanents are given by
\begin{widetext}
\begin{equation}
    \ket{n_1, n_2, \ldots, n_M; t} = \frac{1}{\sqrt{n_1! n_2! \ldots n_M!}} \left( \hat{c}_1^\dagger(t) \right)^{n_1} \cdot \left( \hat{c}_2^\dagger(t) \right)^{n_2} \cdot \ldots \left( \hat{c}_M^\dagger(t) \right)^{n_M} \ket{\rm{vac}}
\end{equation}
\end{widetext}
with $\hat{c}_j^\dagger(t)$ denoting the creation operator of the $j$-th orbital $\phi_j(t)$, 
$\left[ \hat{c}_i(t), \hat{c}_j^\dagger(t) \right] = \delta_{ij} $.
Note that both the coefficients as well as orbitals are time-dependent. Thus, we have to find equations of motions for both
the coefficients and the orbitals. 

As outlined in \cite{B2}, one can employ the Dirac-Frenkel \cite{Var1,Var2},
McLachlan \cite{B23} or Lagrangian variational principle
to derive the (ML-)MCTDHB equations of motion since all variational parameters 
of the ansatz (\ref{EqMCTDHBAnsatz}) are
complex implying the equivalence of these three variational principles \cite{B30}.
Here we will only review the McLachlan approach: For a known wavefunction $\Psi(t)$ at time $t$ the wavefunction at time $t+\tau$ is 
assumed to 
be $\Psi(t+\tau) = \Psi(t) - \frac{i}{\hbar} \theta(t) \tau$, with $\tau$ being small.
The minimum principle for an approximate solution then reads
\begin{equation}\label{mclachlan}
	\vert \vert  \theta -  \hat{H} \Psi \vert \vert^2 = \rm{Min.}	\, ,
\end{equation}
where $\theta$ has to be varied as allowed in view of (\ref{EqMCTDHBAnsatz}) and eventually fixes the optimal value of $\dot{\Psi}$.
Please note that the McLachlan variational principle (\ref{mclachlan}) actually
defines what we mean with the statement that \mbox{(ML-)MCTDHB} provides the with respect to
the ansatz class (\ref{EqMCTDHBAnsatz}) variationally
optimal approximation to the true many-body wave function. In particular,
the (ML-)MCTDHB solutions are variationally optimal with respect to the given
number of orbitals, $M$. Therefore, $M$ has to be increased until the quantities
of interest (observable and their expectation values) become independent of it. In this case, full convergence is achieved.
Explicitly carrying out the variation (\ref{mclachlan}) leads to a coupled system of
ordinary differential equations for the time-dependent coefficients $\boldsymbol{C}(t)$ and
non-linear partial integro-differential equations for the time-dependent orbitals $\ket{\phi_j(t)}$, the MCTDHB equations 
of motion \cite{B17, B4, B18}.

In this work we have employed our implementation of ML-MCTDHB, a generalization of the MCTDHB method, to solve these equations of motion \cite{B5a,B5b}.
The idea beyond ML-MCTDHB is to integrate the ansatz (\ref{EqMCTDHBAnsatz}) in a multi-layer ansatz  \cite{ B3a, B3, B3b} for covering more general scenarios: In such
a multi-layer expansion, strongly correlated degrees of freedom are grouped together and treated as subsystems, which mutually couple to each other. The
resulting cascade in the wave function expansion allows for adapting the ansatz, i.e. the numbers of various basis states, to intra-subsystem and inter-subsystem specific correlations, leading to
a softer scaling of the computational effort. With such a grouping of degrees of freedom, different bosonic species and bosonic atoms in higher-dimensional traps can be simulated more efficiently than by directly applying MCTDHB - if
the correlations between the respective subsystems do not become too strong. For a single bosonic species in a one-dimensional trap, however, there are no distinguished subsytems. Therefore,
the ML-MCTDHB ansatz breaks down to the MCTDHB expansion described above. We refer the reader to \cite{B5a,B5b} for further details on the method.  
Besides, we note that we have employed a harmonic discrete variable representation for representing the orbitals $\ket{\phi_j(t)}$ \cite{D1}.
Concerning the computational costs for obtaining e.g. the contour plot \ref{FigContourFit}, the effort strongly  depends
on the data point $(g,N)$ and the number of orbitals $M$, ranging from a few days up to two weeks (for $g=0.8$, $N=135$, $M=3$) CPU time on
a Intel\textregistered$\,$ Xeon\textregistered$\,$  CPU
E5530  with 2.40GHz. This leads to a total CPU time of the order of eight years for obtaining that plot.

%\newpage
\section{Remarks on the mixing of the center-of-mass and relative motion within MCTDH-type methods}
\label{sec:Appendix_Separation}

In our analysis in section \ref{sec:two-body-breathing-and-beating-dynamics} we have related one of the two 
frequency peaks arising in the breathing mode spectrum to the center-of-mass motion of the system.
Thus, this peak is expected to be independent of the interaction strength.
Please note that in order to make use of the bosonic symmetry in terms of a second quantization representation 
(ML-)MCTDHB solves the time-dependent Schr\"odinger equation in the lab frame, i.e. does not employ the separation of the CM coordinate from the $N-1$ relative coordinates.
Indeed, for a sufficiently high orbital number, we have found that it remains 
almost constant when varying the interaction strength. 
For a low orbital number, however, it tends to move to a higher frequency with 
increasing interaction strength. 

Comparing our results to the standard MCTDH implementation as by the Heidelberg package \cite{B14}, we have seen that both 
methods encounter the same difficulties, namely that this center-of-mass motion peak tends to depend on the interaction 
strength for too low orbitals numbers. Both numerically as well as analytically, we have found evidence that this problem can 
be traced back to an unphysical coupling of center-of-mass and relative motion introduced by truncating the Hilbert space.
The problem is suppressed as mentioned above for large enough orbital numbers. Moreover, it has weak impact on the relative motion spectrum
we are particularly interested in and can be filtered out of the relative motion data. Though, it is quite a subtle convergence issue since it cannot be 
properly detected by estimating the degree of convergence of the MCTDH run by the occupations numbers 
of the highest orbitals as it is commonly done in MCTDH methods.

%\newpage
\bibliography{literatur}
% \bibliography{apssamp}
% \bibliographystyle{unsrt}

\end{document}